\documentclass[
aps,
prb,
onecolumn,
10pt,
superscriptaddress,
nofootinbib,
longbibliography
]{revtex4-2}

\usepackage[T1]{fontenc}
\usepackage{textcomp}
\usepackage{graphicx} 
\usepackage{amsmath}
\usepackage{amsfonts}
\usepackage{enumitem}
\usepackage{braket}
\usepackage{nicematrix}
\usepackage{ragged2e}
\usepackage{lmodern}
\usepackage[dvipsnames]{xcolor}
\usepackage{hyperref}

\hypersetup{
    colorlinks=true,
    linkcolor=black,
    citecolor=NavyBlue,
    urlcolor=RoyalBlue
}

\usepackage{glossaries}
\glsdisablehyper
\newacronym{rsoc}{RSOC}{Rashba spin-orbit coupling}
\newacronym{mbs}{MBS}{Majorana bound state}
\newacronym{mp}{MP}{Majorana polarization}
\newacronym{bdg}{BdG}{Bogoliubov-de Gennes}
\newacronym{sc}{SC}{superconductor}
\newacronym{ssh}{SSH}{Su-Schrieffer-Heeger}
\newacronym{bz}{BZ}{Brillouin zone}
\newacronym{twod}{2D}{two-dimensional}
\newacronym{oned}{1D}{one-dimensional}
\newacronym{zerod}{0D}{zero-dimensional}
\newacronym{ldos}{LDOS}{local density of states}

\begin{document}

\justifying

\newcommand{\yb}[1]{\textcolor{purple}{#1}}
\newcommand{\ang}{\AA}
\newcommand{\sgx}{\mathrm{sgn}(V_x \alpha k_x)}
\newcommand{\sgy}{\mathrm{sgn}(M_{1y} V_y)}
\newcommand{\sgxy}{\mathrm{sgn}(M_{1y} V_y V_x \alpha k_x)}
\newcommand{\sx}{V_x k_x^2}
\newcommand{\sy}{V_y k_y}
\newcommand{\rxx}{\alpha V_x k_x}
\newcommand{\rzx}{\alpha M_{1x} k_x}
\newcommand{\rzy}{\alpha M_{1y} k_y}
\newcommand{\ryy}{\alpha (V_y / 2)}

\title{Corner Majorana states in semi-Dirac materials}

\author{Marta Garc\'{\i}a Olmos}
\affiliation{Nanotechnology Group, USAL--Nanolab and IUFFyM, University of Salamanca, Salamanca, Spain}
\affiliation{Instituto de Estructura de la Materia IEM-CSIC, Serrano 123, Madrid, Spain}

\author{Yuriko Baba}
\affiliation{Instituto de Estructura de la Materia IEM-CSIC, Serrano 123, Madrid, Spain}

\author{Mario Amado}
\affiliation{Nanotechnology Group, USAL--Nanolab and IUFFyM, University of Salamanca, Salamanca, Spain}

\author{Rafael A. Molina}
\affiliation{Instituto de Estructura de la Materia IEM-CSIC, Serrano 123, Madrid, Spain}


\begin{abstract}
    Proximity-induced superconductivity in low-dimensional systems offers a powerful pathway to engineer topological superconducting phases in, otherwise, non-superconducting systems. These exotic phases are of fundamental and technological interest due to the presence of  robust zero-energy modes, the Majorana bound states. In this work, we propose a theoretical framework to realize Majorana bound states from the edge states of a two-dimensional semi-Dirac system. This anisotropic system, under specific conditions, can host non-chiral edge states that propagate only along particular edges, effectively forming separated one-dimensional channels.
    We show that the interplay between Rashba spin–orbit coupling and a Zeeman field on this setup provides the right conditions to get an effective $p$-wave pairing between the edge states by proximity with a $s$-wave superconductor. In finite geometries, each edge can independently undergo a topological phase transition into a one-dimensional topological superconductor and give rise to four zero-energy modes localized at the strip corners. At low energies, the edge states subspace admits a description in terms of coupled Kitaev chains, providing a clear picture of the origin, robustness, and tunability of the corner Majorana modes. Our results establish semi-Dirac materials as a natural platform for realizing Majorana modes in two dimensions without relying on engineered nanostructures, vortices, or crystalline higher-order topology.
\end{abstract}

\maketitle

\section{Introduction}

Topological superconductors host \glspl{mbs}, zero-energy quasiparticles that obey non-Abelian exchange statistics and constitute a promising building block for fault-tolerant quantum computation \cite{Kitaev2003,HasanKane2010, Alicea2012, Sato2017,Ivanov2001, Nayak2008, Sarma2015, Marra2022}. The paradigmatic theoretical example is the one-dimensional Kitaev chain, where spinless fermions with $p$-wave pairing support localized Majorana zero modes at their ends \cite{Kitaev2001}. Realizing such effective spinless $p$-wave superconductivity in solid-state systems, however, requires engineering mechanisms that lift spin degeneracy and convert conventional $s$-wave pairing into an effective odd-parity pairing channel.

\par A practical route, first proposed in semiconductor-superconductor heterostructures \cite{Oreg2010,Lutchyn2010}, relies on the interplay of \gls{rsoc}, Zeeman splitting, and proximity induced $s$-wave pairing. When the chemical potential lies within the Zeeman gap, the pairing projected onto the remaining helical band acquires an effective $p$-wave character, driving the system into a one-dimensional topological superconducting phase. This mechanism has motivated extensive experimental efforts in hybrid nanowires \cite{Stanescu2011,Tewari2012,Prada2020}, but despite some progress \cite{Mourik2012,NadjPerge2014,Deng2016}, the unambiguous identification of topological protected Majorana modes remains challenging after more than a decade of research \cite{Kouwenhoven2025}. Key experimental challenges include disorder-induced segmentation that creates trivial Andreev bound states mimicking Majorana signatures, the required interface quality, and finite-size effects \cite{Liu2012,Prada2012,Aguado2017,DasSarma2023,Prada2020,Karoliya2025}\\[7 pt]

These challenges have motivated the exploration of alternative two-dimensional platforms. One possible route involves embedding quasi-\gls{oned} topological channels within a \gls{twod} bulk, such as in planar Josephson junctions \cite{Hell2017,Pientka2017} as well as topological insulator-superconductor heterostructures following the proposal of Fu and Kane \cite{Fu2008,Fu2009}. By doing so, these systems offer transverse degrees of freedom and enhanced control to mitigate boundary disorder. More recently, the advent of higher-order topological superconductors has provided a distinct route to achieve localized 0D Majorana modes directly at the corners of a \gls{twod} sample, without the need for vortex structures or spatial obstructions \cite{Langbehn2017, Yan2018, Zhang2019, Tan2022}.\\[7 pt]

In this work, we propose that two-dimensional semi-Dirac materials provide a platform where spatially separated one-dimensional channels arise naturally within a higher-dimensional setting. Semi-Dirac systems are characterized by an anisotropic bulk dispersion that is quadratic along one momentum direction and linear along the perpendicular direction \cite{Dietl2008, Goerbig2008, Pardo2009, Banerjee2009, Montambaux2009, Delplace2010, Wu2014, Elsayed2025, Shao2024}. It has been recently shown that, in the band-inverted regime, this anisotropy gives rise to non-chiral edge states that propagate only along specific boundaries of the sample, while remaining absent on others \cite{GarciaOlmos2024,GarciaOlmos2025}. As a consequence, a finite strip geometry hosts spatially separated \gls{oned} edge channels embedded in a \gls{twod} bulk.

Here we demonstrate that, when \gls{rsoc}, a Zeeman field, and conventional $s$-wave pairing are incorporated within a Bogoliubov–de Gennes framework, these semi-Dirac edge states can be driven into a topological superconducting phase. Importantly, the anisotropic nature of the edge spectrum allows each boundary to behave as an effectively independent \gls{oned} channel. We show that the proximity-induced pairing projected onto the edge-state subspace acquires an odd-parity momentum dependence, enabling a mapping onto coupled Kitaev chains localized at opposite edges. In the regime where the chemical potential lies below the Zeeman energy, each edge realizes a non-trivial \gls{oned} topological superconductor, giving rise to four zero-energy Majorana modes localized at the corners of a finite sample.

Our proposal extends the nanowire paradigm to an intrinsically \gls{twod} system, where the spatial separation of edge channels is not engineered lithographically but naturally emerges from the intrinsic anisotropy of the semi-Dirac dispersion. This provides a complementary route to realizing Majorana modes without relying on vortices, magnetic textures, or crystalline higher-order topology.

The paper is organized as follows. In Sec. \ref{sec:model_edge_states}, we introduce the \gls{bdg}  Hamiltonian and describe the electronic edge states of a semi-infinite semi-Dirac system in the absence of superconducting pairing but considering \gls{rsoc} and Zeeman interactions. Using degenerate perturbation theory, we derive the dispersion relation and wave functions of the edge states that are going to be coupled by the superconducting pairing.  We keep the chemical potential and the Zeeman energy inside the bulk gap such that the low-energy physics is governed entirely by the edge states. In Sec. \ref{sec:low_energy_ham} we show that the \gls{bdg} Hamiltonian projected onto the edge states subspace can be decomposed into decoupled sectors that describes the lower and upper boundary and the conventional pairing becomes effectively odd in momentum. Section \ref{sec:Kitaev_mapping} shows how the topological regime in which we find zero-energy modes can be interpreted in terms of coupled Kitaev chains along the $y$-edges of the system. We use the Majorana polarization as a diagnostic beyond spatial localization. Finally, in Sec. \ref{sec:disorder_simulations} we introduce Anderson disorder to the model and study the spatial location of the corner modes, the topological gap closing and the breaking of the degeneracy of the zero-energy modes as a function of the disorder magnitude.

\section{Model and edge states} \label{sec:model_edge_states}

We consider a semi-Dirac system with \gls{rsoc}, subject to a perpendicular Zeeman field and proximitized to an $s$-wave \gls{sc}. This system is described by the following \gls{bdg} Hamiltonian,

\begin{equation}
    \begin{aligned}
        \mathcal{H}(\boldsymbol{k})_{\rm BdG} 
        =&{} - \mu (\tau_z s_0 \sigma_0) + M_{\boldsymbol{k}} (\tau_z s_0 \sigma_z)  
        + V_x k_x^2 (\tau_z s_0 \sigma_x) 
        + V_y k_y (\tau_z s_0 \sigma_y) \\
        &+ 
        \alpha \big[ V_x k_x(\tau_z s_y \sigma_x) 
        - M_{1x} k_x(\tau_z s_y \sigma_z) + M_{1y} k_y(\tau_0 s_x \sigma_z) - \frac{V_y}{2}(\tau_0 s_x \sigma_y) \big] \\
        &+ B_Z (\tau_z s_z \sigma_0) - \Delta (\tau_y s_y \sigma_0) ,
    \end{aligned}
    \label{eq:BdG_Ham}
\end{equation}
where $\sigma_i$, $s_i$ and $\tau_i$ with $i = 0, x,y,z$ denote the identity and Pauli matrices acting on the orbital, spin and particle-hole subspaces, respectively.
The first line of Eq.~\eqref{eq:BdG_Ham}  describes the bare semi-Dirac system, with a mass term given by $M_{\boldsymbol{k}} = M_0 - M_{1x} 
k_x^2 - M_{1y}k_y^2$ and chemical potential $\mu$. The real parameters $V_x$ and $V_y$ govern the quadratic and linear dispersions along the $k_x$ and $k_y$ directions, respectively. The second line corresponds to the \gls{rsoc} contribution, whose strength is set by $\alpha$. Finally, the third line accounts for the Zeeman splitting induced by an out-of-plane magnetic field $B_Z$ and the momentum-independent superconducting pairing amplitude $\Delta$.\\[7 pt] 

For the spin-degenerate limit of the Hamiltonian, obtained by setting $\alpha = B_Z = \Delta = 0$, the system reduces to two identical copies of the bare semi-Dirac model, one for each spin component. In this limit, the mass term $M_{\boldsymbol{k}}$ drives a phase transition between a trivial insulating phase ($M_0/M_{1y} <0$) and a band-inverted regime ($M_0/M_{1y}>0$). 
In the latter regime, in spite of the lack of a \gls{twod} topological invariant, the system hosts quadratic edge states that propagate just along the $x$-direction and remain localized along the $y$-direction \cite{GarciaOlmos2024}. Throughout this work, we focus on this band-inverted regime, keeping both the chemical potential and the Zeeman energy within the bulk gap. This specific parameter space ensures that these edge modes do not hybridize with the bulk states and, consequently, the low-energy physics is dictated entirely by these isolated edge modes. Before introducing the effects of the proximitized \gls{sc}, we first analyze how the spin-dependent interactions modify these states.

\subsection{Edge states in the absence of superconductivity} \label{sec:spin-orbit_edgestates}

We consider a semi-infinite size system with boundaries along the $y$-direction and translational symmetry along the $x$-direction. In this configuration, the system supports quadratic edge states whose dispersion relation is given by, 
\begin{equation}
    E_{0,\xi} = \xi V_x \text{sgn} (M_{1y} V_y) k_x^2,~~~~~\xi = \pm 1.
\end{equation}
These two solutions labeled by $\xi$ are eigenstates of $\sigma_x$ and are exponentially localized at the top or bottom edge according to the sign of $\xi$. Even though they are not chiral, at $k_x=0 \text{ \AA}$, the Hamiltonian reduces to an effective one-dimensional \gls{ssh} model characterized by a non-trivial Zak phase ($\mathcal{Z} = \pi$), which identifies the $k_x = 0\text{ \AA}$ state as topological. Further details can be found in Ref.~\cite{GarciaOlmos2024} and Appendix~\ref{app:details_Hbulk}.
\\[7pt] 

\begin{figure}
    \centering
    \includegraphics[width=0.85\linewidth]{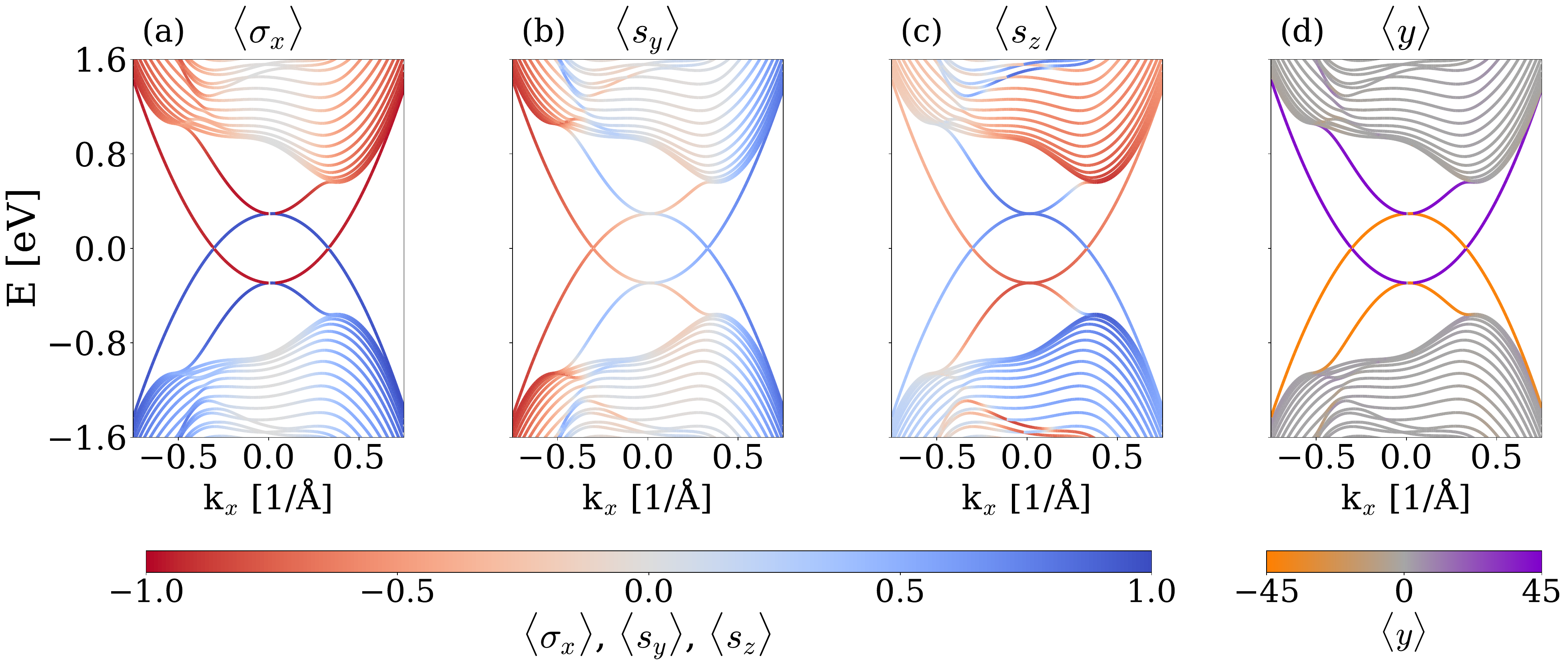}
    \caption{Electronic band structure of a semi-infinite system in $y$-direction (width = 90a) with translational symmetry in $x$-direction. The color encodes the expectation value of the relevant spin and orbital components: (a) $x$-component on the orbital operator, (b) $y$-component and (c) $z$-component of the spin operator, respectively. The rest of the components are nearly zero for the edge states. Panel (d) is colored according to spatial localization in $y$-direction. The color scales range from $-1$ to $+1$ for spin/orbital projections (from red to blue) and from -W/2 to +W/2 for $\langle y \rangle$ (from orange to purple). The edge states remain approximately polarized in the orbital $\sigma_x$ sector, while their spin texture is the competition between the momentum-dependent $s_y$ component induced by the \gls{rsoc} and the $s_z$ component induced by the Zeeman. In particular, the sign of $\langle \sigma_x \rangle$ is correlated with the edge where the state is localized. Parameters are given in units of the bulk gap $M_0$, $M_0 = 1.0 \text{ eV}, M_{1x} = 2.3 \text{ eV}\cdot\text{\AA}{}^{2}, M_{1y} = 2.3 \text{ eV}\cdot\text{\AA}{}^{2}, V_x = -3.8\text{ eV}\cdot\text{\AA}{}^{2}, V_y = -5.0 \text{ eV}\cdot\text{\AA}, \alpha = 0.25 \text{ \AA}{}^{-1}, a = 1 \text{ \AA}, B_Z = 0.35\text{ eV}$. The anisotropy of the spectrum around $k_x=0$ originates from the anisotropy of the Rashba spin-orbit interaction.}
    \label{fig:H4dof_spin_orbital}
\end{figure}

When \gls{rsoc} and Zeeman coupling are included, the SOC lifts the spin degeneracy by inducing momentum-dependent spin mixing, while the Zeeman term produces an additional energy splitting along the $s_z$ direction. As a result, the original spin-degenerate quadratic spectrum evolves into four distinct edge modes where spin is no longer a good quantum number (see Fig. \ref{fig:H4dof_spin_orbital}). To obtain the edge states wave functions, we treat these interactions as perturbations acting within the degenerate subspace of each branch $\xi$, spanned by $\lbrace \ket{\uparrow_z} \otimes \ket{\phi_{\xi}}, \ket{\downarrow_z} \otimes \ket{\phi_{\xi}} \rbrace$. 
The resulting states remain localized in $y$-direction and extended along the edge and can be written as $\ket{\Psi_{\xi, s}(k_x; y)} = \mathcal{N} e^{ik_x x} f_{\xi}(y) \, |e_{\xi, S}(k_x)\rangle$, where $S=\pm 1$ labels the two spin-split branches within a given edge sector $\xi$ and should not be interpreted as a conserved spin quantum number. The orbital spinor $\ket{\phi_{\xi}}$, as well as spatial dependence given by $f_{\xi}(y)$ can be found in Appendix \ref{app:details_Hbulk} and the inner spin-orbital structure denoted by $\ket{e_{\xi,S}}$ is
\begin{align}
    \ket{e_{\xi, +1} (k_x)} &= \left( \cos \frac{\theta_{k_x}}{2} \ket{\uparrow_z} + e^{-i \eta_{\xi}} \sin \frac{\theta_{k_x}}{2} \ket{\downarrow_z} \right) \otimes \ket{\phi_{\xi}}, ~~~~~~ E_{\xi, +1} = E_{0, \xi} + \varepsilon (k_x) ~,\nonumber \\
    \ket{e_{\xi, -1} (k_x)} &= \left( \sin \frac{\theta_{k_x}}{2} \ket{\uparrow_z} - e^{-i \eta_{\xi}} \cos \frac{\theta_{k_x}}{2} \ket{\downarrow_z} \right) \otimes \ket{\phi_{\xi}}, ~~~~~~ E_{\xi, -1} = E_{0, \xi} - \varepsilon (k_x)~,
    \label{eq:spin_evecs}
\end{align}
with
\begin{equation}
    \varepsilon (k_x) = \sqrt{B_Z^2 + (\alpha V_x k_x)^2}, \quad
\cos\theta_{k_x} = \frac{B_Z}{\varepsilon (k_x)}, \quad
\sin\theta_{k_x} = \frac{|\alpha V_x k_x|}{\varepsilon (k_x)},
\quad \eta_\xi = - \frac{\pi}{2} \xi \mathrm{sgn}(M_{1y}V_y \alpha V_x k_x). \nonumber
\end{equation}

Figure~\ref{fig:H4dof_spin_orbital} shows the numerical edge-state spectrum, color-coded according to the expectation values of (a) orbital $\sigma_x$-component, (b) spin $s_y$-component, (c) spin $s_z$-component, and (d) localization in $y$-direction. The remaining orbital and spin components are negligible.
The numerical simulations are performed using a tight-binding regularization of the model on a square lattice with lattice constant $a=1$ and a strip geometry of width $W =90a$, that preserves translational symmetry along $x$. This setup ensures that \gls{oned} edge modes localized on opposite boundaries remain exponentially decoupled.

Figure \ref{fig:H4dof_spin_orbital} show that the orbital character is largely preserved, with the edge states remaining close to eigenstates of $\sigma_x$. By contrast, the spin texture is strongly modified by the combined action of \gls{rsoc} and Zeeman field. This behavior can be understood in terms of an effective two-level description for each branch $H_{\rm spin} = \varepsilon \boldsymbol{n_{\xi}} \cdot \boldsymbol{s}$ with $\boldsymbol{n}_{\xi} = (\sin \theta_{k_x} \cos \eta_{\xi}, - \sin \theta_{k_x} \sin \eta_{\xi}, \cos \theta_{k_x})$. Upon substituting the explicit expressions for $\theta_{k_x}$ and $\eta_{\xi}$, the spin expectation value is confined to the $s_y$-$s_z$ plane, 
\begin{equation}
    \langle \boldsymbol{s}_{\xi, S} \rangle = \frac{1}{\varepsilon (k_x)}\left( 0, \xi \alpha V_x k_x~\text{sgn}(M_{1y} V_y), S B_Z \right)~.
\end{equation}
The Rashba coupling generates an odd $\langle s_y \rangle$ component in $k_x$ and the Zeeman field induces a finite $\langle s_z \rangle$. As a result, the spin quantization axis acquires a momentum dependence along each edge branch, in good agreement with numerical results.

Throughout this section we restrict ourselves to the case $\mu = \Delta = 0$ eV, such that only the electronic sector has been discussed. Although these states are not helical in the strict sense, this momentum-odd spin texture plays an analogous role to the helicity in Rashba nanowires \cite{Das2012}  and suggests that, upon introducing superconducting pairing, the projected edge theory naturally acquires the ingredients required for an effective low-energy \gls{bdg} description with an unconventional pairing structure. 
Within the \gls{bdg} formalism, each electronic eigenstate $\ket{\psi_e(\boldsymbol{k})}$ with energy $E_e(\boldsymbol{k})- \mu$ is accompanied by a hole partner  $\ket{\psi_h (\boldsymbol{k})} = C \ket{\psi_e (-\boldsymbol{k})}$ where $C = \tau_x s_0 \sigma_0 \mathcal{K}$, and with energy $E_h (\boldsymbol{k}) = \mu - E_e(-\boldsymbol{k})$ as required by particle-hole symmetry.

\section{Effective low-energy Hamiltonian} \label{sec:low_energy_ham}

To understand the effect of the induced superconductivity in the edge states, we project the full \gls{bdg} Hamiltonian \eqref{eq:BdG_Ham} onto the low-energy subspace spanned by the electron- and hole-like edge states derived in the previous section, see Eq. \eqref{eq:spin_evecs}. A detailed derivation of this projection is given in Appendix~\ref{app:low-energy_BdG}.

In the limit where the sample width is much larger than the transverse decay lengths ($W \gg \lambda_{1,2}^{-1}$), the edge states localized at opposite edges becomes effectively decoupled. For the normal electronic sector, $\mathcal{H}_{\rm e}$, this follows from the exponential suppression of finite-size hybridization as $W$ increases. In the superconducting sector, $\Delta_{\rm eff}$, the decoupling is enforced by symmetry: the conventional pairing potential acts trivially in the orbital subspace (is proportional to $ \sigma_0$), while the orbital spinors of opposite edges are strictly orthogonal, $\bra{\phi_{+1}} \sigma_0 \ket{\phi_{-1}} = 0$. Consequently, the  full system reduces to two independent one-dimensional \gls{bdg} effective Hamiltonians, $\mathcal{H}_{+1}(k_x)$ and $\mathcal{H}_{-1}(k_x)$, describing each isolated edge. 

Focusing on one edge $\xi$, the relevant low-energy subspace contains two electronic modes and their hole partners. In the Nambu basis $\{ \ket{e_{\xi, +1}}, \ket{e_{\xi, -1}}, \ket{h_{\xi, +1}}, \ket{h_{\xi, -1}} \}$, the effective Hamiltonian takes the form,
\begin{equation} \label{eq:hamBdgxi}
    \mathcal{H}_{\xi} (k_x) = \begin{pmatrix}
h_{\xi}(k_x) & \Delta_{\rm eff, \xi}(k_x) \\ \Delta_{\rm eff, \xi}^{\dagger}(k_x) & -h_{\xi}^*(-k_x)
\end{pmatrix}.
\end{equation}
The normal electronic sector $h_{\xi}(k_x)$ is,
\begin{equation}\label{eq:hameff_xi}
    h_{\xi} = \begin{pmatrix}
        -\mu + \xi V_x k_x^2 \text{sgn}(M_{1y}V_y) + \frac{B_Z^2 - (\alpha V_x k_x)^2}{\varepsilon(k_x)} & 2 \frac{B_Z \alpha V_x k_x}{\varepsilon(k_x)} \\ 2 \frac{B_Z \alpha V_x k_x}{\varepsilon(k_x)} & -\mu + \xi V_x k_x^2 \text{sgn}(M_{1y}V_y) - \frac{B_Z^2 - (\alpha V_x k_x)^2}{\varepsilon(k_x)}
    \end{pmatrix}~,
\end{equation}
and the induced \gls{sc} pairing $\Delta_{\rm eff, \xi} (k_x)$ is given by,
\begin{equation}
    \Delta_{\rm eff, \xi} (k_x)=  -i \xi \frac{\Delta}{\varepsilon(k_x)} \text{sgn}(M_{1y} V_y) \begin{pmatrix}
         \alpha V_x k_x &  -B_Z \text{sgn}( \alpha V_x k_x) \\  -B_Z \text{sgn}(\alpha V_x k_x) & -\alpha V_x k_x & 
    \end{pmatrix}~.
    \label{eq:hameff_Delta}
\end{equation}

\begin{figure}
    \centering
    \includegraphics[width=0.78\linewidth]{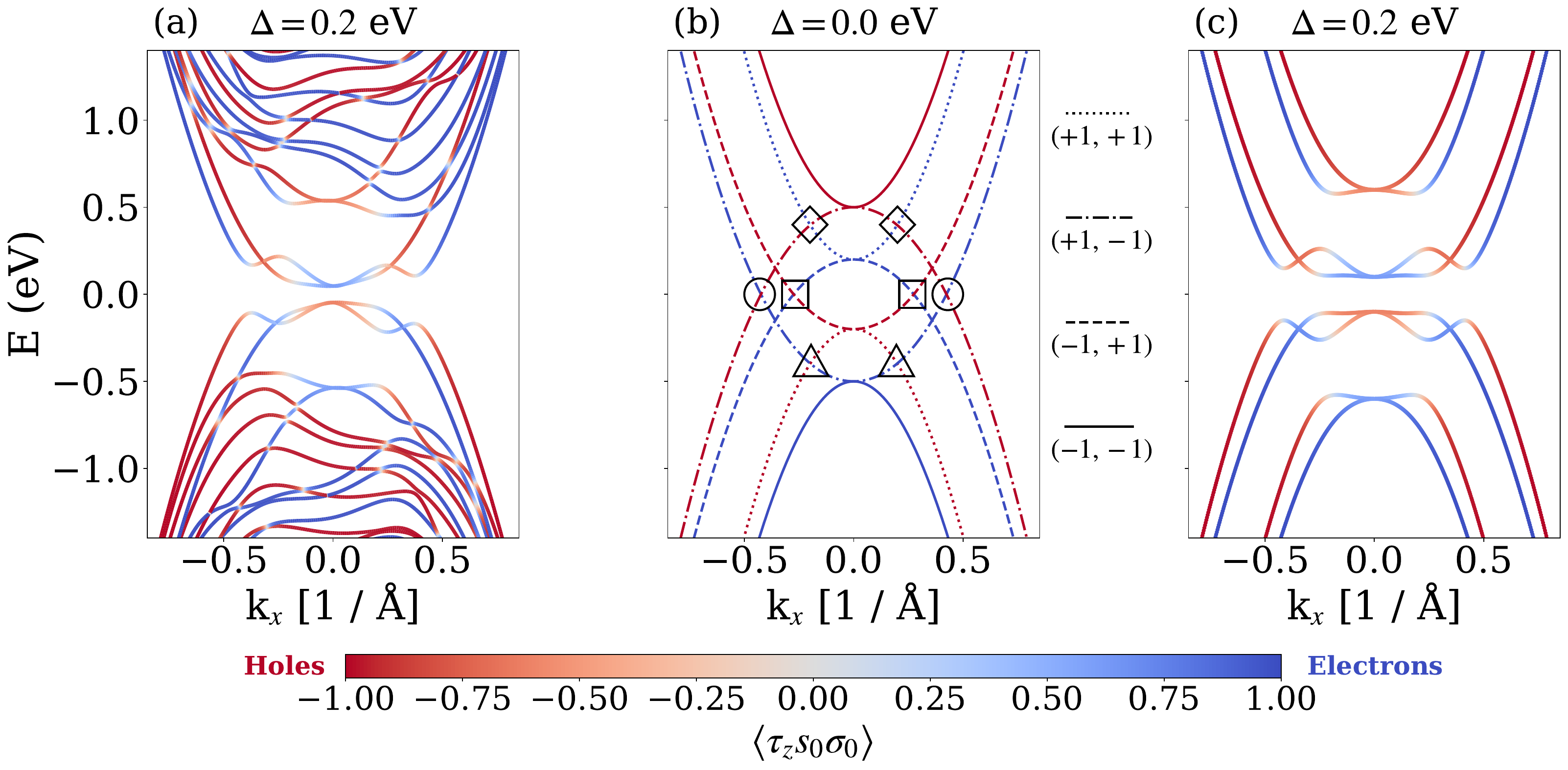}
    \caption{Band dispersion of the semi-infinite system in the $y$-direction with translational symmetry in the $x$-direction in Nambu space. Panel (a) shows the numerical spectrum with $\Delta = 0.2$ eV of the \gls{bdg} Hamiltonian of Eq.~\eqref{eq:BdG_Ham} discretized in a strip with translational symmetry in the $x$-direction and a finite width of $W=90a$ in the $y$-direction. In panel (b) we plot the analytical energies obtained perturbatively given by Eq.~\eqref{eq:spin_evecs} for $\mu = 0.15$ eV for electrons and $\mu = -0.15$ eV for holes, in the absence of the superconducting coupling. We use the line style to differentiate the distinct branches $(\xi, S)$: a dotted line for $(\xi, S) = (+1,+1)$, dotted-dashed line for $(\xi, S) = (+1,-1)$, a dashed line for $(\xi, S) = (-1,+1)$ and a continuum line for $(\xi, S) = (-1,-1)$. The markers (diamonds, circles, squares and triangles) highlight the couplings between edge states induced by superconductivity, as derived in Eq.~\eqref{eq:effective_SCcoupling}. Panel (c) shows the band dispersion obtained from the low-energy Hamiltonian display in Eq.~\eqref{eq:low-energy_BdG} for $\Delta = 0.2$ eV. 
    Red (blue) indicates hole-like (electron-like) states in the \gls{bdg} sense. The parameters are the same as in Fig.~\ref{fig:H4dof_spin_orbital}.}
    \label{fig:low_energy_spectrum}
\end{figure}

Figure \ref{fig:low_energy_spectrum} compares the spectra obtained from the full Hamiltonian~\eqref{eq:BdG_Ham} in a nanoribbon geometry and the effective low-energy Hamiltonian~\eqref{eq:hamBdgxi}. 
Panel (a) shows the numerical spectrum of the full Hamiltonian for a nanoribbon finite in $y$-direction and panel (c) displays the spectrum of the effective low-energy theory. 
In panel (b) we include the analytical dispersion relation of Eq.~\ref{eq:spin_evecs} for electrons ($\mu > 0$) and holes ($\mu < 0$) without superconducting coupling, $\Delta = 0$, in order to highlight the good agreement between the expected gap openings according to non-zero elements of $\Delta_{\rm eff, \xi} (k_x)$ and the numerical results. We use the color scale to show the electron (in blue) or hole (in red) character of the states.\\[7 pt]

To understand this figure, we first discuss the induced pairing structure. $\Delta_{\rm eff, \xi} (k_x)$, determines which electron-hole pairs can hybridize. The diagonal elements of this matrix couple an electronic state, $\ket{e_{\xi, S}}$ and its hole partner $\ket{h_{\xi, S}}$, these terms are strictly odd in momentum, and control the gap openings at Fermi level. By contrast, the off-diagonal elements couple $\ket{e_{\xi, S}}$ to $\ket{h_{\xi, S'}}$. Their amplitude is governed by the Zeeman energy but retains an odd character encoded in the factor $\text{sgn}(\alpha V_x k_x)$. These inter-branch pairing, with $S' \neq S$, generates additional avoided crossing away from the Fermi level. 
For the parameter regime considered in Fig. \ref{fig:low_energy_spectrum}, $B_Z > \mu > 0$, the states coupled at Fermi level are the pair
$\ket{e_{+1, -1}}-\ket{h_{+1, -1}}$ and the pair $\ket{e_{-1, +1}}-\ket{h_{-1, +1}}$, 
shown as circles and squares in panel (b), respectively. At higher and lower energies, the ones that coincide energetically and can hybridize are  $\ket{e_{+1, +1}}$ with $\ket{h_{+1, -1}}$ (indicated with diamonds) and $\ket{e_{+1, -1}}$ with $\ket{h_{+1, +1}}$ (indicated with triangles). Notably, the $\text{sgn}(k_x)$ dependence originates from the phase parametrization of edge spinors in Eq.~\eqref{eq:spin_evecs}. The \gls{rsoc} within the degenerate subspace $\lbrace \ket{\uparrow_z} \otimes \ket{\phi_{\xi}}, \ket{\downarrow_z} \otimes \ket{\phi_{\xi}}\rbrace$ defines an effective two-level Hamiltonian whose off-diagonal elements are proportional to $\alpha V_x k_x$. Writing this complex spin mixing in the polar form introduces the phase $\eta_{\xi}(k_x)$ that absorbs the sign of $k_x$ and some matrix elements the low-energy Hamiltonian acquire explicit $\text{sgn}(k_x)$ factors. \\[7 pt]

Therefore, although superconductivity is introduced through a conventional spin-singlet pairing term, which 
originally couples purely spin-up and spin-down electrons, its projection onto the edge state subspace leads to a qualitatively different pairing structure. As we have mentioned before, the origin of this effective odd-parity pairing lies in the spin structure of the edge states. Although the purely electronic edge states lack strict helicity, each branch $\ket{e_{\xi, \pm 1}}$ is a momentum-dependent superposition of spin-up and spin-down configurations. The interplay of \gls{rsoc} and the Zeeman field twists this spin texture, successfully projecting the uniform $s$-wave gap into an effective odd-parity pairing with respect to $k_x$. 

\section{Kitaev chains mapping and Majorana zero modes} \label{sec:Kitaev_mapping}

\begin{figure}
    \centering
    \includegraphics[width=0.6\linewidth]{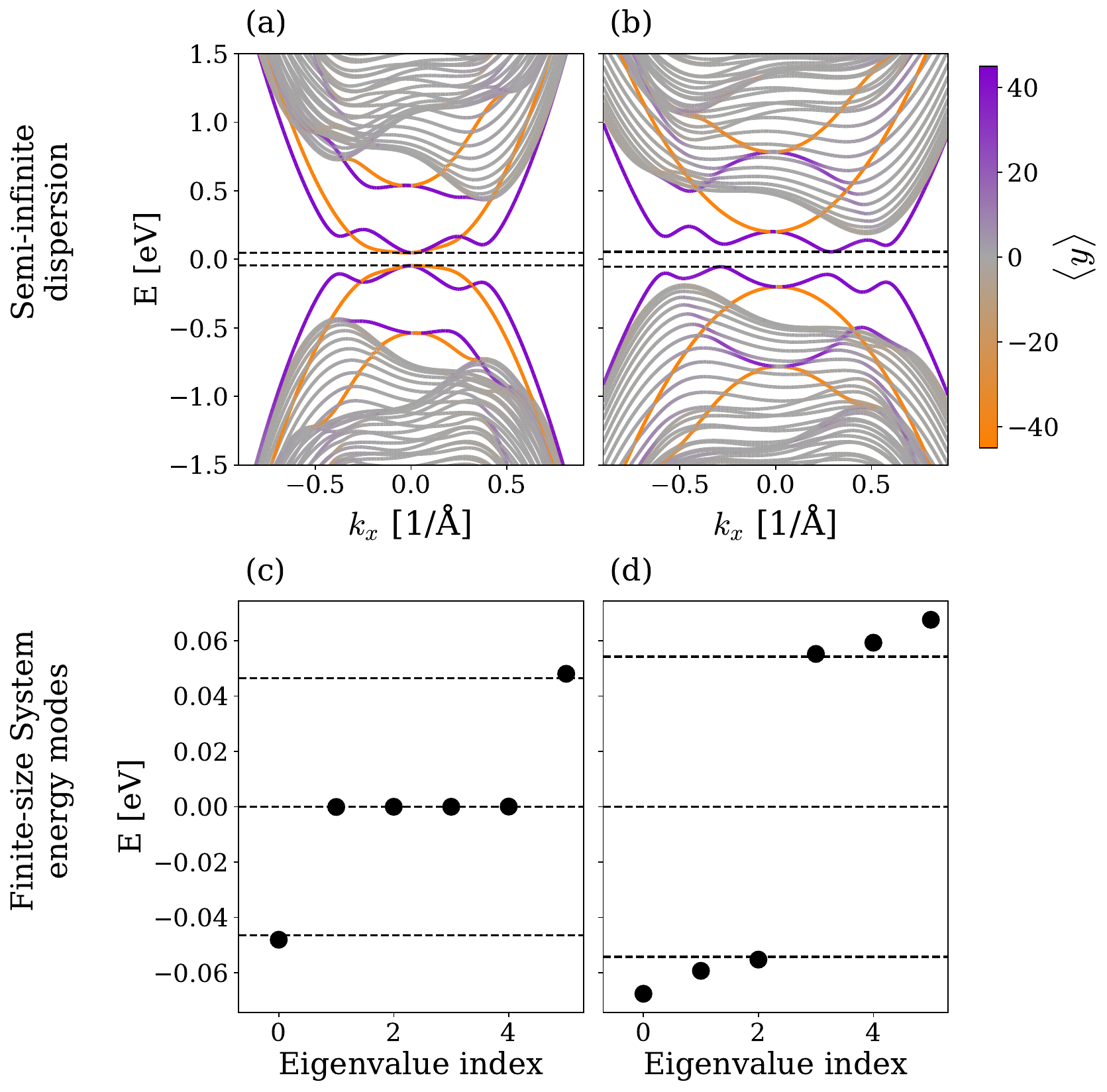}
    \caption{Panels (a) and (b) display the bands of a system with translational invariance along $x$-direction and $W = 90a$. The color scale indicates the expectation value of the y-coordinate. Panels (c) and (d) display the corresponding energy spectrum of the finite-size system in both directions. Panels (a) and (c) correspond to the regime with $\mu < B_Z$, we take $\mu = 0.15 \text{ eV}$  and $B_Z = 0.35 \text{ eV}$, where we find zero-energy modes inside the effective \gls{sc} gap plotted with dashed lines. In contrast, panels (b) and (d) show the results for the regime without Majorana bound states, in which the low-energy coupled states belong to the same edge, i.e. $\mu > B_Z$, with $\mu = 0.45 \text{ eV}$, $B_Z = 0.35 \text{ eV}$. The rest of the model parameters are the ones of Fig. \ref{fig:low_energy_spectrum}}
    \label{fig:mu_vs_Bz}.
\end{figure}

The topological consequences of this dynamically generated odd-parity pairing depend fundamentally on the placement of the chemical potential $\mu$ relative to the Zeeman gap. We can distinguish two different regimes. For $|\mu| > |B_Z|$, the chemical potential intersects both spin-orbit branches of a given edge (upper edge states if $\mu > 0$ and lower edge states if $\mu < 0$) while for $|\mu| < |B_Z|$ the states that cross the Fermi level belong to different edges, leaving only a single propagating mode per edge at the Fermi level where spin is not a good quantum number. In this limit, the odd-parity elements of ${\Delta}_{\xi} (k_x)$ fulfill the essential prerequisites for a topological \gls{sc} phase \cite{Oreg2010, Lutchyn2010}. Figure \ref{fig:mu_vs_Bz} shows the numerical spectrum of the full \gls{bdg} Hamiltonian in a semi-infinite system, highlighting the localization of the edge states in the transverse direction, $y$. In panel (a), with $B_Z > \mu > 0$, the system is in the topological regime, and in panel (b), the system is in the trivial regime with $\mu > B_Z > 0$.\\[7 pt]

To observe the topological zero-energy signatures, we study the states that appear on a finite-size system of width $W =90a$, and length $L = 180a$. In the trivial regime, the spectrum remains fully gapped [see Fig. \ref{fig:mu_vs_Bz} (d)] and, tuning $\mu$ into the topological regime, four zero-energy modes emerge inside the effective superconducting gap [see Fig. \ref{fig:mu_vs_Bz} (c)].
These states are found to be localized at the corners of the sample and to be topologically protected, as discussed in the following. 
\\[7 pt] 

To evaluate rigorously the topological phase boundaries and characterize these modes analytically, we note that the Zeeman splitting breaks time-reversal symmetry, placing the system in the Altland-Zirnbauer symmetry class D. In one dimension, the topology of class D is characterized by a $\mathbb{Z}_2$ Pfaffian invariant $\nu$ \cite{Ryu2010, Schnyder2008, Chiu2016}. This invariant is determined by the sign of the ratio of the Pfaffians evaluated strictly at the time-reversal invariant momenta (TRIM) of the Brillouin zone ($k_x = 0, \pi$) \cite{Fu2007, Wimmer2012}:
\begin{equation} \label{eq:pfaffian}
    \nu = \text{sgn} \left( \frac{\rm Pf [A (0)]}{\rm Pf[A(\pi)]} \right),
\end{equation}
where $A$ is a real antisymmetric matrix constructed by using the particle-hole symmetry operator, $A(k_x) = \mathcal{H}(k_x) \tau_x$.\\[7 pt]

To compute this invariant, we regularized the continuum effective Hamiltonian onto a \gls{oned} lattice by making the standard substitution, $k_i \rightarrow \frac{1}{a} \sin (k_i a)$, $k_i^2 \rightarrow \frac{2}{a^2} \cos(k_i a)$. Taking the lattice constant $a = 1 \text{ \AA}$, the periodic \gls{oned} Hamiltonian for a given edge $\xi$ takes the structure of two coupled chains. If we momentarily isolate a single spin-orbit branch $S$, the effective model describes a single-channel superconducting wire, 
\begin{equation}
    \mathcal{H}_{{\xi, S}} (k_x) = d_z(k_x) \tau_z + d_y (k_x) \tau_y 
\end{equation}
where $\tau_i$ acts in the electron-hole subspace and
\begin{align}
    d_z (k_x) &=  -\mu + S \frac{B_Z^2}{\varepsilon(k_x)} + 2 \left( \xi V_x \text{sgn}(M_{1y} V_y) - S\frac{(\alpha V_x)^2}{\varepsilon(k_x)} \right) [1-\cos(k_x)], \\
    d_y (k_x) &= (\xi S)\Delta \text{sgn}(M_{1y} V_y) \frac{\alpha V_x \sin(k_x)}{\varepsilon (k_x)},
\end{align}
with the regularized dispersion $\varepsilon(k_x) = \sqrt{B_Z^2+ 4(\alpha V_x)^2 \sin^2(k_x/2)}$. Note that we replace $|k_x| \rightarrow \frac{2}{a}\sin(k_xa/2)$ to maintain proper parity and prevent additional inversions of the bands near $k_x = \pi$.\\[7 pt]

The former Hamiltonian resembles to a Kitaev chain \cite{Kitaev2001} enabling us to interpret the effective low-energy Hamiltonian of the full system as if each edge $\xi$ supports two coupled Kitaev-like chains, corresponding to the two branches $S = \pm 1$. 
In fact, the edge Hamiltonian $\mathcal{H}_{\xi}$ is not merely two independent Kitaev chains; it contains off-diagonal kinetic and pairing terms that explicitly couple the $S = +1$ and $S = -1$ branches. These are the off-diagonal contribution given in Eq.~\eqref{eq:hameff_xi} and \eqref{eq:hameff_Delta}. However, because these inter-branch couplings are proportional to odd functions of momentum, they vanish at $k_x = 0$ and $k_x = \pi$ and at these specific TRIM points the $4\times4$ edge Hamiltonian perfectly block-diagonalizes into the two uncoupled $S$ sectors per edge. The global topological invariant of the entire system is then identical to the product of the Pfaffians of the individual Kitaev chains,
\begin{equation}
    \nu = \prod_{S = \pm 1 , \xi = \pm 1} \nu_{S, \xi}~,  \qquad \nu_{S, \xi} = \text{sgn}\big[ \big(- \mu + S|B_Z| \big) \big(-\mu + S |B_Z| + 4  \xi V_x \text{sgn}(M_{1y} V_y) \big) \big]
\end{equation}

\begin{figure}[htb]
    \centering
    \includegraphics[width=0.9\linewidth]{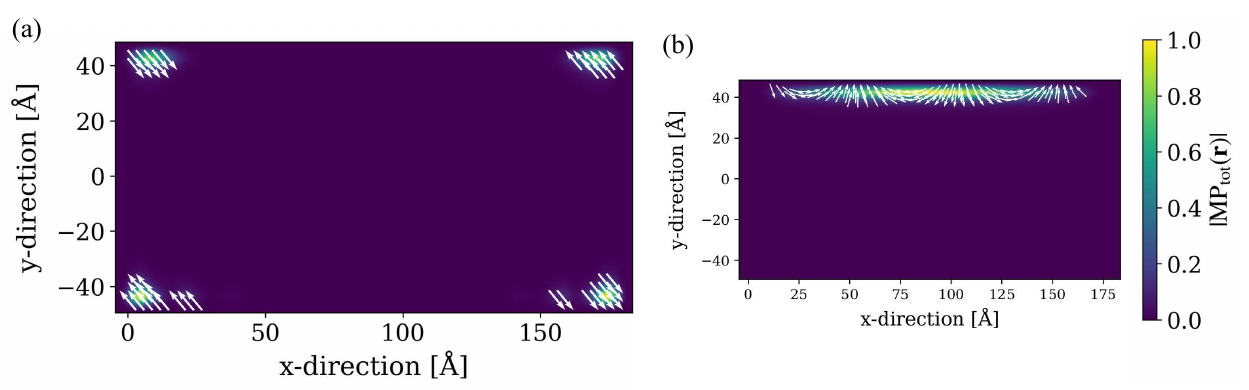}
    \caption{ Spatially resolved Majorana polarization (\gls{mp}). The
    arrows indicate the real and imaginary part of \gls{mp} in each site. The colormap corresponds to its absolute value. In (a) we consider the four zero-energy states calculated in the maximally localized basis by diagonalizing the position operator, $\hat{X}+\hat{Y}$ projected onto that subspace. The uniform alignment of the \gls{mp} vectors indicates a well-defined Majorana phase for the corner modes. (b) shows the \gls{mp} for the first state above the superconducting gap, which is an extended state along the upper edge. Unlike the zero-energy modes, the precession of the arrows indicates trivial behavior.}
    \label{fig:LDOS_Majoranas}
\end{figure}

This quantity always gives $+1$, because for any value of $\mu$ and the Zeeman energy, two of the four Kitaev chains undergo a band inversion. However, the physical consequences of these inversions depend entirely on where those inverted chains reside spatially. Since the opposite edges of the sample are exponentially decoupled ($W \gg \lambda_{1,2}^{-1}$), an inverted chain on the upper edge cannot couple to and annihilate with an inverted chain on the lower edge. We can then evaluate the invariant for each edge independently $\nu_\xi = (\nu_{\xi, +1}, \nu_{\xi, -1})$. On the table \ref{tab:regimes} we summarize the different options that one can find. In the trivial regime one of the edges has both chains  inverted, while on the opposite edge none of them are inverted. In the topological regime, each edge exhibits an inverted chain while the other remains trivial. Each edge realize a non-trivial \gls{oned} topological \gls{sc}, supporting zero-energy modes at its boundaries, i.e. at the corners of the \gls{twod} sample, as shown in Fig. \ref{fig:LDOS_Majoranas}.\\[7 pt]

\begin{table}[ht]
\centering
\begin{tabular}{c | c | c | c}
\hline
Regime &
$\boldsymbol{\nu}_{\xi,S}
=
(\nu_{+1,+1},\,\nu_{+1,-1},\,\nu_{-1,+1},\,\nu_{-1,-1})$
&
$(\nu_{\xi=+1},\,\nu_{\xi=-1})$
 & Phase \\
\hline
$\mu > B_Z > 0$
&
$(-1,-1,+1,+1)$
&
$(+1,+1)$
&
Trivial \\
$B_Z > \mu > 0$
&
$(+1,-1,-1,+1)$
&
$(-1,-1)$
& 
Topological\\
$0 > \mu > B_Z$
&
$(+1,-1,-1,+1)$
&
$(-1,-1)$
& 
Topological
\\
$0 > B_Z > \mu$
&
$(+1,+1,-1,-1)$
&
$(+1,+1)$
& Trivial
\\
\hline
\end{tabular}
\caption{Classification of the topological phases in terms of the relation between the chemical potential and the Zeeman field. The second column lists the Pfaffians $\nu_{\xi,S}$ of each Kitaev chains and the third one corresponds to the Pfaffian associated with each sector $\xi$. While the global invariant is always $\nu=+1$, different spatial distributions of the inverted chains distinguish the trivial phase, with both inversions on the same edge, from the topological phase, with one inversion on each edge.}
\label{tab:regimes}
\end{table}

To verify the Majorana character of the zero-energy modes $\ket{\psi_i}$, we compute the site-resolved Majorana polarization, defined as the local expectation value of the particle-hole operator $C = \tau_x s_0 \sigma_0  \mathcal{K}$, 
%
\begin{equation}
    \text{MP}_i(\boldsymbol{r}) = \bra{\psi_i} C \hat{r}\ket{\psi_i}~,
\end{equation}
where $\hat{r} = \ket{r} \bra{r}$ projects onto the site $r$. This complex quantity can be viewed as the local contribution of the particle-hole overlap of the BdG eigenstate: its magnitude measures the strength of the local electron-hole coherence, while its phase encodes the relative phase between the electron and hole components. To visualize the full zero-energy subspace, we sum the contributions of the four zero modes, $\text{MP}_{\rm tot}(\boldsymbol{r}) = \sum_{i=1}^4 \text{MP}_i(r)$ and represent $|\text{MP}_{\rm tot}(\boldsymbol{r})|$ as a density plot in Fig. ~\ref{fig:LDOS_Majoranas}. The white arrows indicate the local orientation of the $\text{MP}_{\rm tot}$.

A useful integrated diagnostic is
\begin{equation}
    C_R^{(i)} = \frac{ \left| \sum_{\hat{r} \in \mathcal{R}} \bra{\psi_i} C \hat{r} \ket{\psi_i} \right|}{\sum_{\hat{r} \in \mathcal{R}} \bra{\psi_i} \hat{r} \ket{\psi_i}},
\end{equation}
where $\mathcal{R}$ is the spatial region in which the state is localized. For an ideal Majorana  bound state, one expects $C_R^{(i)} \approx 1$. This is a necessary condition but not sufficient, since values close to unity may also occur for Andreev bound states (ABSs) \cite{SedlmayrBena2015, Karoliya2025}. However, in contrast with the ABSs, for a MBS the local particle-hole overlap is expected to remain phase-coherent over the region where the state is localized. 
 
In our numerical calculations, we find $C_{R}^{(i)}$ to be extremely close to unity for the corner states. Consistently, Fig.~\ref{fig:LDOS_Majoranas} (a) shows that within each corner the MP vectors are uniformly aligned and pointing in opposite directions for the modes localized on the same edge, which is the expected behavior for a MBS. For comparison, Fig. \ref{fig:LDOS_Majoranas} (b) shows the Majorana polarization of the first localized state with energy $E = 0.05\text{ eV}$, that is extended along the upper edge. In that case, the spatial precessing pattern reveals a non-Majorana character. 

\begin{figure}[h]
    \centering
    \includegraphics[width=0.6\linewidth]{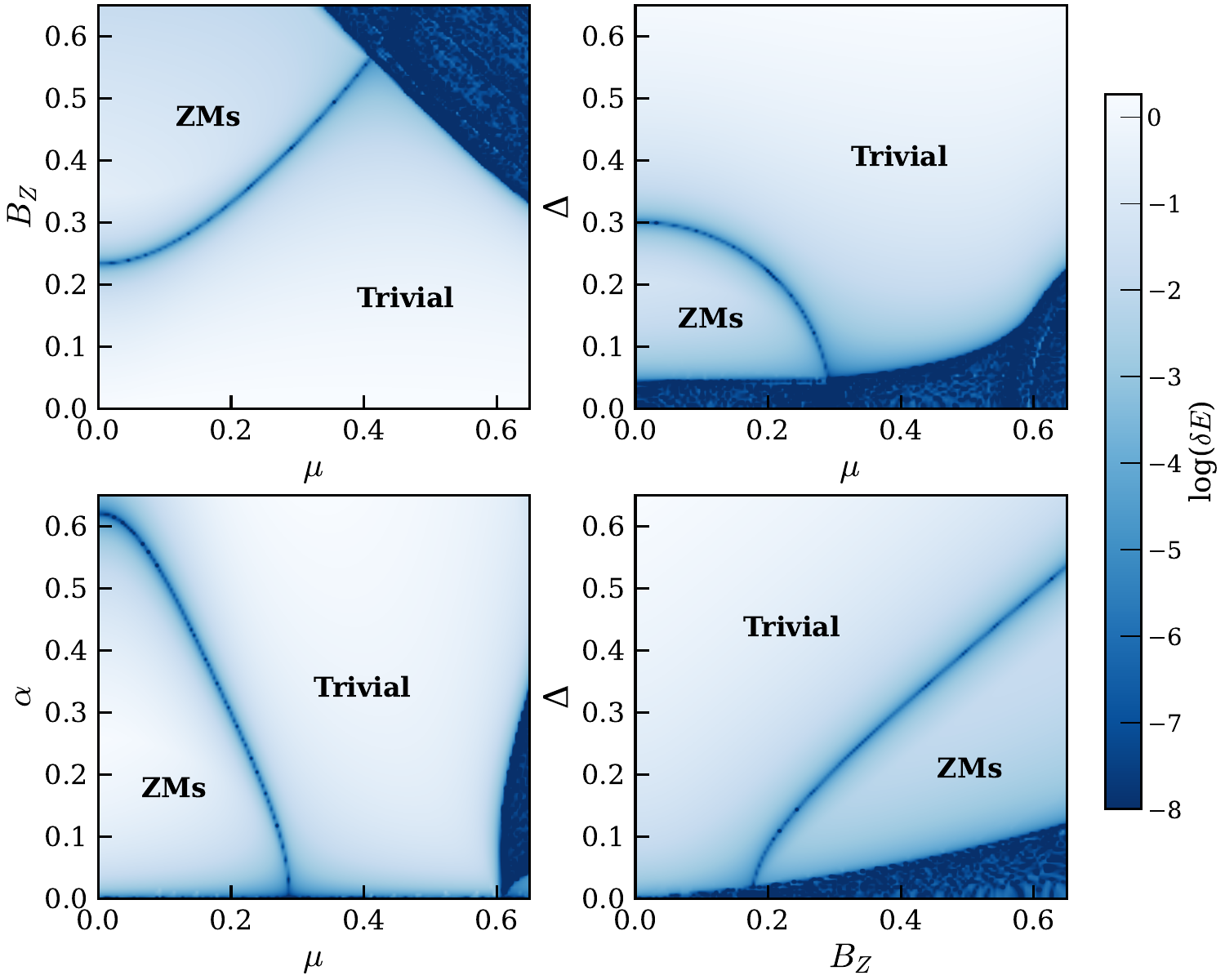}
    \caption{Gap size $\delta E$ as a function of the model parameters for the semi-infinite system with translational symmetry along $x$-direction and $W = 90~\text{ \AA}$. Dark blue indicates gap closings, plotted on a logarithmic color scale. The phase boundaries between a topological phase with Majorana zero-energy modes labeled by ZMs and a trivial one, are determined by gap closings of the edge states spectrum at $k_x = 0 \text{ \AA}$. We consider the same set of parameters as in Fig.~\ref{fig:H4dof_spin_orbital} varying the ones indicated in the axis.}
    \label{fig:gapmap_closings}
\end{figure}

In Fig. \ref{fig:gapmap_closings} we display the gap magnitude of the semi-infinite spectrum for \gls{twod} cuts of the parameter space. We have identified two phases separated by gap closings at $k_x = 0 \text{ \AA}$, shown in dark blue. The region labeled ZMs correspond to the phase that supports zero-energy corner modes in the finite geometry. The remaining region is topologically trivial. 

\section{Resilience to disorder} \label{sec:disorder_simulations}

To quantify the robustness of corner states against disorder, we compute the integrated local density of states (LDOS) in the corner regions and analyze the energy level spacing between adjacent states. 
We have considered Anderson disorder, which is an onsite potential that assigns a random magnitude uniformly distributed in the interval $[-w/2, w/2]$ at a lattice site. We have chosen that this disorder preserves particle-hole symmetry and then enters the \gls{bdg} Hamiltonian as $\epsilon_i \Gamma_{z00}$, where $\epsilon_i$ is the random on-site energy at site $i$. All results are averaged over 200 disorder realizations.
For each disorder realization, we numerically diagonalize the Hamiltonian, extracting the six central eigenvalues around zero energy. 
\\[7 pt]

\begin{figure}
    \centering
    \includegraphics[width=0.9\linewidth]{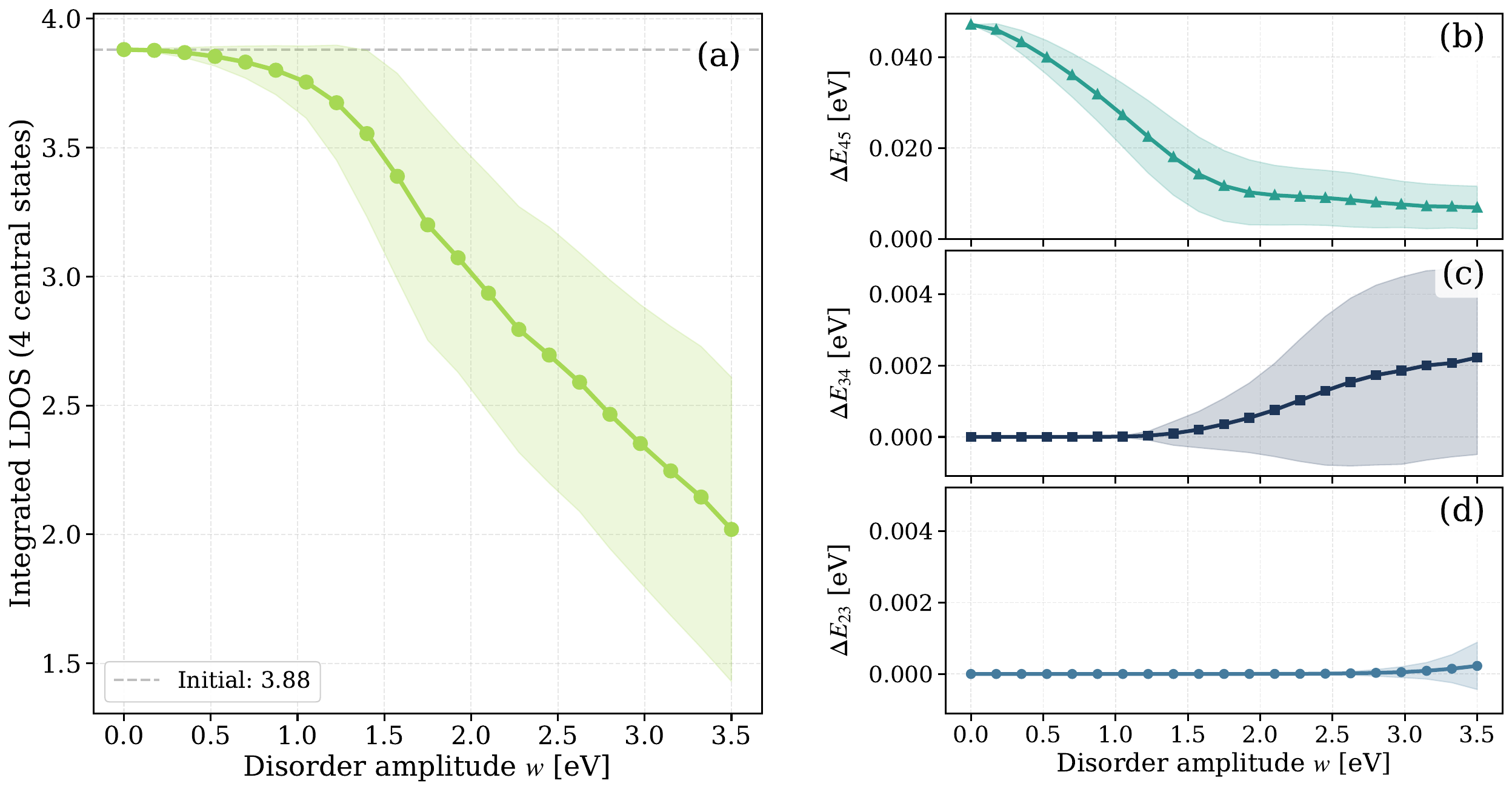}
    \caption{(a) Integrated \gls{ldos} of the four central states within the fixed corner mask as a function of the disorder strength $w$. The dashed horizontal line marks the clean-limit value, 3.88. The right panels (b-d) show the level spacing of consecutive states as a function of the disorder strength $w$. We extract by numerical diagonalization six states around zero energy, in the clean limit, eigenstates 0 and 5 are extended states along the edge, while 1, 2, 3, and 4 are the four corner states. The labels of the gaps follow these indexes. Accordingly, panel (b) shows $\Delta E_{45}$, the effective superconducting gap; panel (c) shows $\Delta E_{34}$, which quantifies the energy splitting between upper and lower corner states, and panel (d) shows $\Delta E_{23}$, the energy difference between two states of the same edge localized in opposite corners. Shaded regions correspond to the standard deviations of the average value over 200 disorder realizations.}
    \label{fig:disorder_ldos_spacing}
\end{figure}

In Fig.~\ref{fig:disorder_ldos_spacing} (a) we show the evolution of the integrated LDOS at the corners as a function of the magnitude of the Anderson disorder $w$. To define the corner region, we first compute the site-resolved \gls{ldos}, $\rho_i$, associated with the four central states in the clean system, summing over the internal degrees of freedom at each site. Sites with $\rho_i < 10^{-4}$ are discarded, and the remaining ones are ranked in descending order of $\rho_i$. The mask is then defined as the minimal set of sites with the highest weight  that captures 97\% of the integrated \gls{ldos}. This yields a mask of 779 sites, corresponding to the 4.73\% of the total system size, concentrated around the four corners. We then track how much \gls{ldos} remains localized in this fixed region as disorder increases. 

In parallel, we analyze the loss of degeneracy at zero energy of the corner modes and their spectral separation from the effective bulk, given in this case by \gls{oned} states along the edges. In the clean limit, states labeled as 1,2,3,4 correspond to the four zero-energy states, while states with indexes 0 and 5 are extended states along $x$-direction.
Hence, we compute three energy gaps $\Delta E_{ij} = |E_j-E_i|$ to
\begin{itemize}
    \item $\Delta E_{23}$: The splitting between two partner states belonging to the same edge.
    \item $\Delta E_{34}$: The splitting between corner states localized at opposite edges.
    \item $\Delta E_{45}$: The gap between the corner zero-energy modes and the first one in the effective bulk.
\end{itemize}
In the clean limit, four corner modes are numerically degenerate at zero energy, with $\Delta E_{23} \sim 10^{-9}$ eV, $\Delta E_{34} \sim 10^{-7}$ eV and a topological gap of $\Delta E_{45} \sim 0.047$ eV. 
Figure \ref{fig:disorder_ldos_spacing} summarizes the evolution of these quantities as a function of the disorder strength. Panel (a) shows the integrated \gls{ldos} of the four central states within the fixed corner mask, while panels (b)–(d) display the disorder-averaged level spacings. In all panels, the shaded regions indicate one standard deviation over 200 disorder realizations, quantifying the fluctuations induced by disorder.\\[7 pt]

A characteristic disorder scale can be identified around $w = 1.5$ eV. Up to this disorder amplitude, the integrated \gls{ldos} remains predominantly localized in the fixed corners. In panel (b), the effective topological gap $\Delta E_{45}$, decreases steadily with disorder but remains clearly finite up to this amplitude and in panel (c) $\Delta E_{34}$, and, above all, its standard deviation, grows progressively with disorder starting at $1.5$ eV.  This indicates that we loose for some configurations the degeneracy between upper and lower states. Since $w = 1.5$ eV is about 32 times larger than the clean effective topological gap, this identifies a remarkably broad window of disorder robustness. Notably, around $w = 2.3$ eV, the standard deviation of $\Delta E_{45}$ and $\Delta E_{34}$ overlap possibly indicating an Anderson transition where the \gls{oned} edge states become highly localized and the 0D states cannot be clearly distinguishable. 

By contrast, panel (d) shows that $\Delta E_{23}$ is substantially less affected by the disorder over the whole range studied. This behavior highlights the independence between the states on the upper edge and those on the lower edge and, in particular, the different decay lengths of the \gls{oned} edge states from which they originate.\\[7 pt]

These results demonstrate that the corner states are remarkably robust, maintaining localization and degeneracy even when disorder magnitude exceeds significantly the effective gap and it is compared with the bulk gap $\sim 1.0$ eV.

\section{Conclusions}

We have demonstrated that semi-Dirac materials provide a natural platform for realizing proximity-induced topological superconductivity and Majorana zero modes in an intrinsically two-dimensional system. The key ingredient is the strong anisotropy of the semi-Dirac dispersion, which gives rise to boundary-selective quadratic edge states that propagate exclusively along specific edges of the sample. This spatial selectivity effectively generates decoupled one-dimensional channels embedded in a two-dimensional bulk.

By introducing \gls{rsoc} and a perpendicular Zeeman field, conventional spin-singlet $s$-wave pairing projected onto the edge-state subspace acquires an odd-parity momentum dependence. The internal spin structure of the edge modes converts this into an effective odd-momentum pairing along each boundary. The resulting low-energy Bogoliubov–de Gennes Hamiltonian can be mapped onto coupled Kitaev chains localized at opposite edges of the system.

Our analysis reveals that the topological phase is governed by the interplay between the chemical potential and the Zeeman energy. When the chemical potential lies below the Zeeman energy, only one effective Kitaev channel per edge becomes topologically non-trivial. Because opposite edges are exponentially decoupled, their contributions do not cancel, and each boundary realizes a one-dimensional topological \gls{sc}. In a finite geometry, this leads to four zero-energy modes localized at the sample corners, which we identify as Majorana bound states through both their topological characterization and their spatially resolved Majorana polarization. In contrast, when the chemical potential exceeds the Zeeman energy, two coupled Kitaev channels per edge become simultaneously inverted. Their hybridization allows the Majorana modes to gap out, yielding a topologically trivial phase. This highlights the crucial role of anisotropy and spatial separation in stabilizing the topological regime.

Importantly, the Majorana corner states found here do not rely on vortices, magnetic textures, or crystalline higher-order topology. Instead, they emerge from the effective one-dimensional nature of the semi-Dirac edge spectrum. In this sense, our proposal extends the nanowire paradigm to a two-dimensional material where the spatial separation of topological channels is intrinsic rather than engineered. Beyond its conceptual interest, this mechanism suggests new avenues for engineering Majorana modes in anisotropic quantum materials and heterostructures where edge-selective transport naturally occurs. 

From an experimental perspective, semi-Dirac dispersions have been predicted or observed in a variety of material platforms, including VO$_2$/TiO$_2$ oxide nanostructures \cite{Pardo2009}, $\alpha$-(BEDT-TTF)$_2$I$_3$ under pressure \cite{Katayama2006}, phosphorene under strain \cite{CastellanosGomez2014,Rodin2014}, thin films of Cd$_3$As$_2$ \cite{Liu2022}, silicene oxide \cite{Zhong2016}, and ZrSiS \cite{Shao2024}. In addition, it has been shown that uniaxial strain applied to graphene, or more generally to a honeycomb lattice, can drive the merging of Dirac cones into a semi-Dirac point \cite{Montambaux2009b}. Particularly appealing is the case of Na$_3$Bi, whose low-energy Hamiltonian along the (100) direction coincides with the semi-Dirac model considered here, including the presence of a band-inversion mass term. This suggests that appropriately oriented thin films or heterostructures based on Dirac semimetals could provide a realistic starting point for implementing our proposal. Combining such materials with proximity-induced superconductivity, \gls{rsoc} (either intrinsic or interface-enhanced), and moderate Zeeman fields may therefore offer a viable route toward realizing the Majorana corner states predicted in this work.

\section*{Acknowledgments}

This work has been supported by Comunidad de Madrid “Talento Program” (Grant 2019-T1/IND-14088), the Agencia Estatal de Investigación from Spain (Grant PID2022-136285NB-C31/C32) and FEDER/Junta de Castilla y León Research (Grant No. SA106P23). M. G. O. acknowledges financial support from the Consejería de Educación, Junta de Castilla y León, and ERDF/FEDER.

\section*{Funding}

This work has been supported by the Agencia Estatal de Investigación from Spain (MCIN/AEI/10.13039/501100011033) under Grant PID2022-136285NB-C31/C32 and FEDER/Junta de Castilla y León Research (Grant No. SA106P23). M. G. O. acknowledges FEDER/Junta de Castilla y León Research Grant No. SA121P20.

\bibliographystyle{apsrev4-2}
\bibliography{Bibliography}

\appendix

\section{Details of the bulk Hamiltonian without superconducting pairing} \label{app:details_Hbulk}

\begin{figure}[h]
    \centering
    \includegraphics[width=0.8\linewidth]{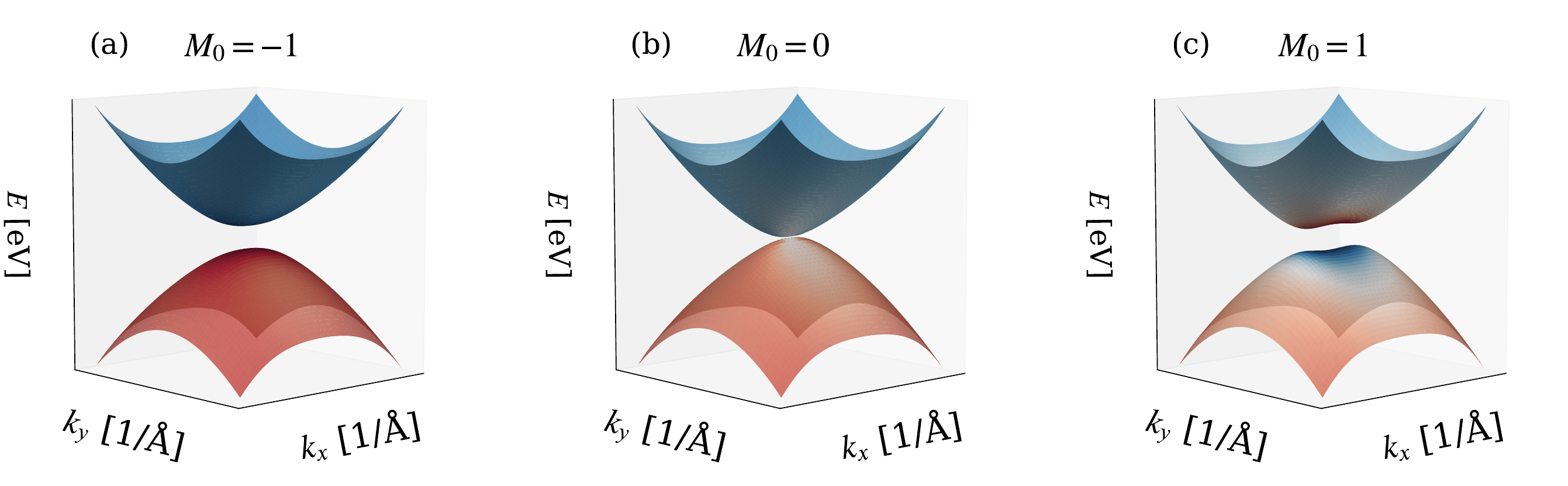}
    \caption{Bulk spectrum of a semi-Dirac system in terms of the sgn($M_0/M_{1y}$). For $M_0 = 0$ eV the gap is closed, distinguishing a phase transition between an insulating phase and a band inverted regime shown with the expectation value of the $\sigma_z$ operator.}
    \label{fig:BB2dof_sigma_z}
\end{figure}

The semi-Dirac spinful Hamiltonian in the absence of any spin-dependent interaction is given by 
\begin{equation}
     \mathcal{H}_{\rm SD} (\boldsymbol{k}) =  s_0 (M_{\boldsymbol{k}} \sigma_z + V_x k_x^2 \sigma_x + V_y k_y \sigma_y) 
\end{equation}
where $s_i$ acts in the spin space and $\sigma_j$ in the orbital space. Importantly, the Hamiltonian for one spin sector is already time-reversal symmetric, which means that when we consider both spin species as time-reversal partners, we get two identical copies of the orbital Hamiltonian studied in Ref. \cite{GarciaOlmos2024}. The characteristic feature of the spinless model is that it exhibits an anisotropic dispersion relation in momentum space, quadratic in $k_x$ and linear in $k_y$. The mass term, $M_{\boldsymbol{k}} = M_0 - M_{1x}k_x^2-M_{1y} k_y^2$, drives a phase transition between a trivial insulating phase and a band-inverted regime that we show in Fig. \ref{fig:BB2dof_sigma_z}.\\

\begin{figure}[h]
    \centering
    \includegraphics[width=0.85\linewidth]{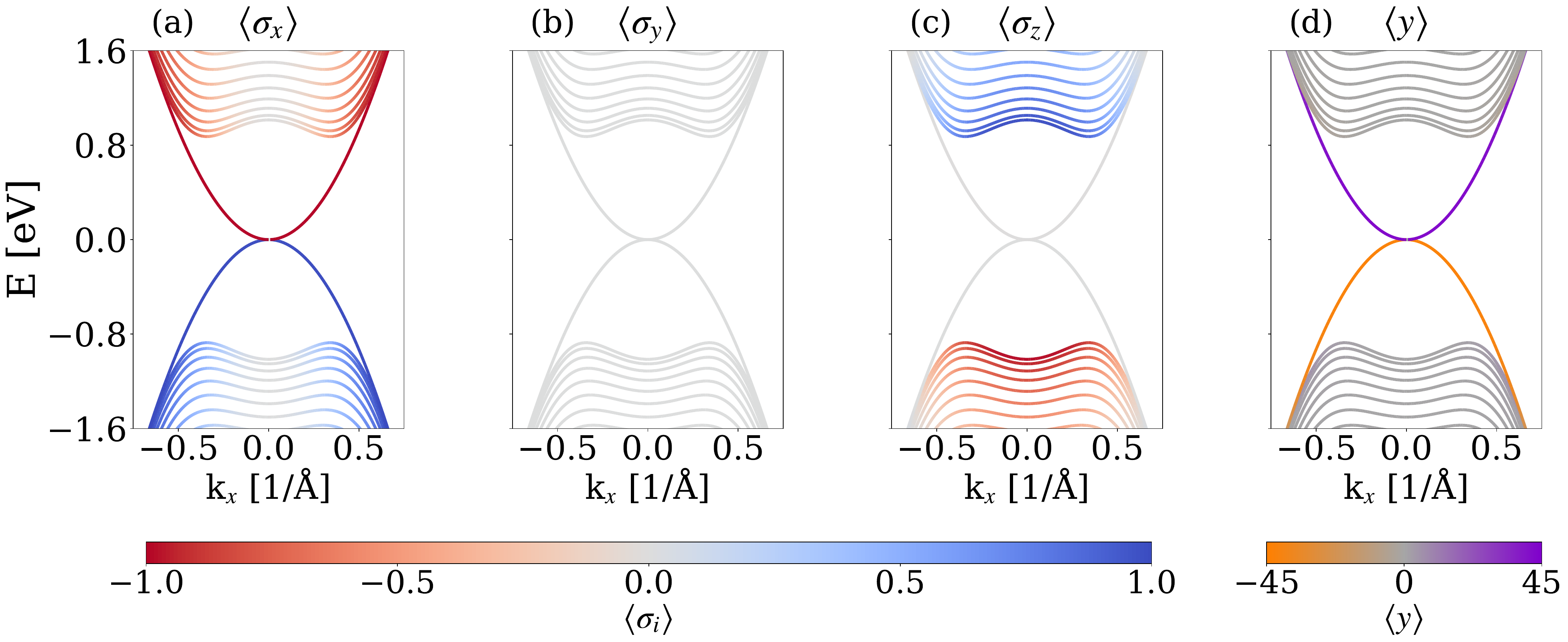}
    \caption{Electronic band structure of a semi-infinite system in $y$-direction (width = 90a) with translational symmetry in $x$-direction. The colormap shows the expectation value of the orbital part (a)-(c) and panel (d) shows the expectation value of the $y$-coordinate. The color scales range from $-1$ to $+1$ (from red to blue) and from -W/2 to +W/2 for spatial location (from orange to purple). The edge states are essentially polarized in the $x$ component, whose value is associated with the edge on which they live.
    This figure has been made using parameters normalized to the bulk gap $M_0$, $\mu = 0.0 \text{ eV}, M_0 = 1.0 \text{ eV}, M_{1x} = 2.3 \text{ eV}\cdot\text{\AA}{}^{2}, M_{1y} = 2.3 \text{ eV}\cdot\text{\AA}{}^{2}, V_x = -3.8\text{ eV}\cdot\text{\AA}{}^{2}, V_y = -5.0 \text{ eV}\cdot\text{\AA}, \alpha = 0.0 \text{ \AA}{}^{-1}, a = 1 \text{ \AA}, B_Z = 0.0\text{ eV})$.}
    \label{fig:2dof_spectrumFINITE}
\end{figure}

To reveal the edge modes that this system hosts, we consider a square lattice with a width of 90a, where $a$ is the lattice parameter, and  translational symmetry in the $x$-direction. In this configuration, $k_x$ remains a good quantum number, while $k_y$ is replaced by the operator $-i\partial_y$. In the topological regime, edge states appear as exponentially localized solutions near the boundaries at $y = \pm W/2$ but no boundary modes arise if the system is instead finite along the $x$-direction. 
These edge states exhibit a quadratic dispersion relation labeled by $\xi = \pm 1$,
\begin{equation}
    E_{0,\xi} = \xi V_x \text{sgn} (M_{1y} V_y) k_x^2,
\end{equation}
The two parabolas, degenerated in spin, have opposite curvature and touch each other at $k_x = 0$, as shown in Fig. \ref{fig:2dof_spectrumFINITE}. 
In real space, these solutions are given by,
\begin{equation}
    \boldsymbol{\psi}_{\xi}(k_x, y) = \mathcal{N}e^{ik_x x} f_{\xi}(y) \phi_{\xi}
\end{equation}
where the spatial dependence, $f_{\xi}(y)$ and the orbital spinor $\phi_{\xi}$ are given by 
\begin{align}
    f_{\xi}( y) &= \left[e^{\lambda_1(\xi y - W/2)} - e^{\lambda_2(\xi y - W/2)}\right]  \label{eq:spatial_func}\\
    \phi_{\xi} &= \frac{1}{\sqrt{2}} \begin{pmatrix} 1 \\ \xi \,\text{sgn}(M_{1y}V_y)\end{pmatrix} \label{eq:orbital_spinor}
\end{align}
and the decay lengths are 
\begin{align} 
    \lambda_{1,2} = \frac{V_y}{M_{1y}} \pm \frac{1}{M_{1y}} \sqrt{V_y^2 -M_0 M_{1y} + M_{1x} M_{1y}k_x^2}. \label{eq:decay_lengths}
\end{align}
In terms of these quantities, the normalization constant $\mathcal{N}$ is given by, $|\mathcal{N}|^2 = 2 \lambda_1 \lambda_2 (\lambda_1 + \lambda_2)/(\lambda_1-\lambda_2)^2$.
That means that they are polarized in $\sigma_x$ in the orbital space, and their expectation value determines the edge in which they live. The states with $\langle  \sigma_x \rangle  = -1$ are localized in the upper edge and the ones with $\langle  \sigma_x \rangle  = +1$ are localized in the lower edge. The situation can be inverted by changing the sign of the coupling term, $V_x$. 
This means that they are not associated with a single orbital but instead they live in a bonding or antibonding superposition of the two orbitals A and B.\\[7pt]

In Fig. \ref{fig:2dof_spectrumFINITE} we plot the bands of the semi-infinite system and the expectation value of the orbital components numerically obtained. We have chosen the lattice constant $a = 1 \text{ \AA}$ and an illustrative  set of parameters, $M_0 = 1.0 \text{ eV}, M_{1x} = 2.3 \text{ eV}\cdot\text{\AA}{}^{2}, M_{1y} = 2.3 \text{ eV}\cdot\text{\AA}{}^{2}, V_x = -3.8\text{ eV}\cdot\text{\AA}{}^{2}, V_y = -5.0 \text{ eV}\cdot\text{\AA}, \alpha = 0.25 \text{ \AA}{}^{-1}, a = 1 \text{ \AA}, B_Z = 0.35\text{ eV})$ to keep the edge states well separated from the bulk states in the relevant low-energy window and with enough width to ensure localization of the edges.

When the \gls{rsoc} and the Zeeman splitting are included, the original spin-degenerate quadratic spectrum evolves into four distinct edge modes where spin is no longer a good quantum number. The \gls{rsoc} that arises from structural inversion symmetry breaking, can be induced, for instance, with an external electric field perpendicular to the plane of the sample \cite{Winkler2003}. This electric field is perceived as a magnetic field in the rest frame of the moving carriers, which interacts with their spin, resulting in a momentum-dependent spin mixing that splits the band except for $k_x = 0$. In order to also break this degeneracy point, a Zeeman term is included. As a result, we end up with two bands associated with edge states in the upper edge (positive curvature parabolas), two bands associated with edge states in the lower edge (negative curvature parabolas) with a spin that rotates in the $s_y$-$s_z$ plane. Notably, if we place the chemical potential below the Zeeman energy, each edge has an isolated spinless band, a key ingredient to realize a Kitaev chain. 

To compute the wavefunctions, we apply degenerate perturbation theory within the subspace spanned by $\lbrace \ket{\uparrow_z} \otimes \ket{\phi_{\xi}}, \ket{\downarrow_z} \otimes \ket{\phi_{\xi}} \rbrace$ for each edge sector $\xi$. Thus, we get an effective two-level Hamiltonian,
\begin{equation}
    H_{z, \xi} = \begin{pmatrix}
        B_Z & \chi_{\xi} \\ \chi_{\xi}^* & -B_Z
    \end{pmatrix}
\end{equation}
where $\chi_{\xi} = \xi V_x \text{sgn}(M_{1y}V_y) \alpha k_x$. 

\section{Analytical derivation of the Effective low-energy BdG Hamiltonian} \label{app:low-energy_BdG}

In this section we provide a step-by-step derivation of the full low-energy \gls{bdg} Hamiltonian. This effective model is obtained by projecting the bulk \gls{bdg} Hamiltonian of Eq.~\eqref{eq:BdG_Ham} onto the subspace spanned by the electronic edge states and their respective hole partners. These edge modes are solutions of a semi-infinite system confined between $y= \pm W/2$ with translational symmetry in $x$-direction. Consequently, only $k_x$ remains a good quantum number and the bulk Hamiltonian, $\mathcal{H}_{\rm BdG}(k_x,k_y)$ in this geometry must be evaluated by replacing the transverse momentum with the differential operator, $k_y \rightarrow -i \partial_y$.\\[7 pt]

In the extended particle-hole space, the basis states are eight-component spinors that can be compactly written as, 
\begin{align}
    \ket{\Psi_{\tau, \xi, S} (k_x; y)} = \mathcal{N} e^{i k_x x} f_{\xi}(y)  \ket{\chi_{\tau, \xi, S} (k_x)}, 
\end{align}
where $\mathcal{N}$ is the normalization constant and $f_{\xi}(y)$ is the spatial profile (see Eq.~\eqref{eq:spatial_func}) that describes the exponential decay from each edge, labeled by $\xi \in \lbrace -1,+1 \rbrace$. The spinor $\ket{\chi_{\tau, \xi, S}}$ encodes the inner particle-hole, spin and orbital structure. Specifically, the electronic edge states are given by $\ket{\chi_{e, \xi, S}} = \ket{\tau_+} \otimes \ket{e_{\xi,S}(k_x)}$ where $\tau_+ = (1,0)^T $ and the hole-like states are given by $\ket{\chi_{h, \xi, S}} = \ket{\tau_-} \otimes \ket{e_{\xi, S}(-k_x)}^*$ with $\tau_- = (0,1)^T$. We remember that,
\begin{align}
    \ket{e_{\xi, +1} (k_x)} &= \ket{s_{\xi, +1}(k_x)} \otimes \ket{\phi_{\xi}} = \frac{1}{\sqrt{2}} \left( \cos \frac{\theta_{k_x}}{2} \ket{\uparrow} + e^{-i \eta_{\xi}} \sin \frac{\theta_{k_x}}{2} \ket{\downarrow} \right) \otimes \begin{pmatrix}
        1 \\ \xi \text{sgn}(M_{1y} V_y)
    \end{pmatrix},\\
    \ket{e_{\xi, -1} (k_x)} &= \ket{s_{\xi, -1}(k_x)} \otimes \ket{\phi_{\xi}} = \frac{1}{\sqrt{2}} \left( \sin \frac{\theta_{k_x}}{2} \ket{\uparrow} - e^{-i \eta_{\xi}} \cos \frac{\theta_{k_x}}{2} \ket{\downarrow} \right) \otimes \begin{pmatrix}
        1 \\ \xi \text{sgn}(M_{1y} V_y)
    \end{pmatrix}
\end{align}

Thus, the resulting effective model inherits a  block structure in Nambu space. We only need to compute two independent matrices: the normal electronic block $H_{\rm e}(k_x)$ and the induced pairing $\Delta_{\rm eff} (k_x)$,
\begin{equation}
    \mathcal{H}^{\rm eff}_{\rm BdG} (k_x) = \begin{pmatrix}
        H_{\rm e}(k_x) & \Delta_{\rm eff}(k_x) \\ \Delta_{\rm eff}^{\dagger}(k_x) & H_{\rm h}(k_x)
    \end{pmatrix}
\end{equation}
The corresponding hole block is completely determined by particle-hole symmetry, $H_h(k_x) = -H_e^*(-k_x)$.\\[7 pt]

Each matrix element is computed via the inner product over all degrees of freedom (spatial and inner space),
\begin{equation}
    \left[ \mathcal{H}^{\rm eff} (k_x) \right]_{\tau \xi S, \tau' \xi' S'} = \int dy~ \bra{\Psi_{\tau, \xi, S} (k_x; y)}  \mathcal{H}_{\rm BdG} (k_x, -i\partial_y) \ket{\Psi_{\tau', \xi', S'} (k_x; y)}.
    \label{eq:tensor_projection}
\end{equation}
Because both the bulk Hamiltonian and the wavefunctions of the edge states can be factorized into a tensor product of the inner degrees of freedom contributions, we can systematically perform the calculation of the matrix elements by evaluating each subspace separately.

To guide this procedure, we explicitly recall the bulk \gls{bdg} Hamiltonian from Eq.~\eqref{eq:BdG_Ham}:
\begin{equation}
    \begin{aligned}
        \mathcal{H}(\boldsymbol{k})_{\rm BdG} 
        =&{} - \mu (\tau_z s_0 \sigma_0) + M_{\boldsymbol{k}} (\tau_z s_0 \sigma_z)  
        + V_x k_x^2 (\tau_z s_0 \sigma_x) 
        + V_y k_y (\tau_z s_0 \sigma_y) \\
        &+ 
        \alpha \big[ V_x k_x(\tau_z s_y \sigma_x) 
        - M_{1x} k_x(\tau_z s_y \sigma_z) + M_{1y} k_y(\tau_0 s_x \sigma_z) - \frac{V_y}{2}(\tau_0 s_x \sigma_y) \big] \\
        &+ B_Z (\tau_z s_z \sigma_0) - \Delta (\tau_y s_y \sigma_0) ,
    \end{aligned}
\end{equation}

\noindent \textbf{The pairing block: $\Delta_{\rm eff}(k_x)$}\ 
The coupling between the electron and hole states is mediated exclusively by the conventional superconducting coupling term, $\Delta  (\tau_y s_y \sigma_0)$, as all other terms of the Hamiltonian are diagonal in Nambu space.\\[7 pt] 

This term cannot couple electron and holes states localized at the opposite edges edge. This is because the orbital spinors of opposite edges are orthogonal and the pairing operator acts trivially in the orbital space, $\bra{\phi_{+1}}\sigma_0 \ket{\phi_{-1}} = 0$. Furthermore, since the superconducting pairing is a momentum-independent quantity, the spatial integration over $y$ is straightforward and serves to define the normalization constant, $\mathcal{N}$,
\begin{equation}
    I_0 = |\mathcal{N}|^{-2} = \int_0^{\infty} dz f_{\xi}^*(z) f_{\xi}(z) = \frac{(\lambda_1-\lambda_2)^2}{2\lambda_1 \lambda_2 (\lambda_1 + \lambda_2)}.
\end{equation}
where we used the coordinate $z = W/2 - \xi y$ and the decay lengths $\lambda_{1,2}$ from Eq.~\eqref{eq:decay_lengths}. This ensures that the induced pairing is independent of the size of the system. Evaluating the remaining overlaps the effective pairing is given by,
\begin{equation}
    {\Delta}_{\rm eff} (k_x)=  -i \frac{\Delta}{\varepsilon(k_x)} \text{sgn}(M_{1y} V_y) \begin{pmatrix}
         \alpha V_x k_x &  -B_Z \text{sgn}( \alpha V_x k_x) & 0 & 0 \\  -B_Z \text{sgn}(\alpha V_x k_x) & -\alpha V_x k_x & 0 & 0 \\
        0 & 0 & -\alpha V_x k_x & B_Z \text{sgn}( \alpha V_x k_x) \\ 0 & 0 & B_Z \text{sgn}( \alpha V_x k_x) & \alpha V_x k_x
    \end{pmatrix}
    \label{eq:effective_SCcoupling}
\end{equation}
written in Nambu basis $\lbrace e, h \rbrace$.

\noindent \textbf{Electronic block: $H_{\rm e}(k_x)$}\
The electronic block $H_e (k_x)$ contains a richer structure due to the presence of terms up to second order in $k_y$ (within the mass term $M_{\boldsymbol{k}} = M_0 - M_{1x}k_x^2 - M_{1y}k_y^2$).  We decompose this block into diagonal subblocks representing the intra-edge couplings $h_{\xi}$, and off-diagonal subblocks representing the inter-edge hybridization matrix $\mathcal{T}$:
\begin{equation}
H_{\rm e}(k_x) = \begin{pmatrix}
h_{+1}(k_x) & \mathcal{T}(k_x) \\ \mathcal{T}^{\dagger}(k_x) & h_{-1}(k_x)
\end{pmatrix}, \qquad \text{with} \quad \mathcal{T}(k_x) = \begin{pmatrix}
\mathcal{T}_{+, +} & \mathcal{T}_{+, -} \\ \mathcal{T}_{-, +} & \mathcal{T}_{-, -}
\end{pmatrix}.
\end{equation}
Here, the subscripts in $\mathcal{T}_{S, S'}$ denote the coupling between the spin-orbit branch $S$ at the top edge ($\xi=+1$) and the branch $S'$ at the bottom edge ($\xi=-1$).

\begin{enumerate}[label=\Alph*.]
    \item \textbf{Intra-edge coupling ($\xi = \xi'$)} \\
    For states localized at the same boundary, the projection onto the orbital and spin space simplifies significantly the calculation. In particular, all surviving terms are independent of the transverse momentum $k_y$, so the spatial integration in $y$ becomes trivial.  It simply yields the normalization constant ($I_0 = 1$), removing any need to evaluate derivative operators. After performing the remaining spin-orbital algebra onto the perturbed Rashba-Zeeman basis, the non-zero elements are,
    \begin{equation}
        h_{\xi}(k_x) = \begin{pmatrix}
            -\mu + \xi V_x k_x^2 \text{sgn}(M_{1y}V_y) + \frac{B_Z^2 - (\alpha V_x k_x)^2}{\varepsilon(k_x)} & 2 \frac{B_Z \alpha V_x k_x}{\varepsilon(k_x)} \\ 2 \frac{B_Z \alpha V_x k_x}{\varepsilon(k_x)} & -\mu + \xi V_x k_x^2 \text{sgn}(M_{1y}V_y) - \frac{B_Z^2 - (\alpha V_x k_x)^2}{\varepsilon(k_x)}
        \end{pmatrix}
    \end{equation}
    whose diagonalization returns the expected energy $E_{\xi, S} = -\mu + E_{0,\xi} + S \varepsilon(k_x)$ exactly as expected.
    
    \item \textbf{Inter-edge coupling ($\xi \neq \xi'$)} \\
    In the case of opposite edge states, the spatial overlap requires the evaluation of the three distinct overlap integrals, $I_{\xi \xi'}^n = \int_{-W/2}^{W/2} dy~f_{\xi} (y) (-i \partial_y)^n f_{\xi'}(y)$, for $n = 0,1,2$, which factorize analytically as:
    \begin{align}
        &I_{+-}^0  
        = |\mathcal{N}|^2 \left[ e^{-\lambda_1 W} \left( W + \frac{2}{\lambda_1 - \lambda_2} \right) + e^{-\lambda_2 W} \left( W - \frac{2}{\lambda_1 - \lambda_2} \right) \right] \\
        & I_{+-}^1 
        = i |\mathcal{N}|^2 \left[ e^{-\lambda_1 W} \left( W\lambda_1 + \frac{\lambda_1 + \lambda_2}{\lambda_1 - \lambda_2} \right) + e^{-\lambda_2 W} \left( W\lambda_2 - \frac{\lambda_1 + \lambda_2}{\lambda_1 - \lambda_2} \right) \right] \\
        & I_{+-}^2 
        = -|\mathcal{N}|^2 \left[ e^{-\lambda_1 W} \left( W\lambda_1^2 + \frac{\lambda_1^2 + \lambda_2^2}{\lambda_1 - \lambda_2} \right) + e^{-\lambda_2 W} \left( W\lambda_2^2 - \frac{\lambda_1^2 + \lambda_2^2}{\lambda_1 - \lambda_2} \right) \right] 
    \end{align}
\end{enumerate}
Evaluating the full spin-orbital algebra alongside these spatial integrals, the matrix elements of the inter-edge hybridization $\mathcal{T}(k_x)$ are given by:
\begin{align}
    \mathcal{T}_{S, S} &= S \left[ \frac{(M_0-M_{1x}k_x^2) B_Z + \frac{1}{2}V_y \alpha^2 V_x k_x}{\varepsilon (k_x)}\right] I_{+-}^0 + i \xi \frac{\text{sgn}(M_{1y}V_y)}{\varepsilon (k_x)} \Big( V_y B_Z + M_{1y} \alpha^2 V_x k_x \Big) I_{+-}^1 - S M_{1y} I_{+-}^2, \\
    \mathcal{T}_{S, -S} &= \left[ \frac{(M_0-M_{1x}k_x^2) |\alpha V_x k_x| - \frac{1}{2}V_y \alpha B_Z \text{sgn}(\alpha V_x k_x)}{\varepsilon (k_x)} + S \xi M_{1x} k_x \alpha \text{sgn} (M_{1y} V_y V_x \alpha k_x)\right] I_{+-}^0 \nonumber \\
    &\quad + i \xi \frac{\text{sgn}(M_{1y}V_y)}{\varepsilon (k_x)} \Big( V_y |\alpha V_x k_x| - M_{1y} \alpha B_Z \text{sgn}(\alpha V_x k_x) \Big) I_{+-}^1 - M_{1y} I_{+-}^2.
\end{align}

These exact form transparently demonstrate that all inter-edge coupling terms scale  with the exponential factors $e^{-\lambda_1 W}$ and $e^{-\lambda_2 W}$. Then, when the sample width is large enough ($W \gg \lambda_{1,2}^{-1}$), the off-diagonal matrix elements become entirely negligible and the low-energy physics is well capture by the decoupled structure
\begin{equation}
    H_{\rm e} (k_x) = 
    \left(
    \begin{array}{c|c}
    h_{\xi = +1}(k_x) & 0_{4\times4} \\ \hline
    0_{4\times4} & h_{\xi = -1}(k_x)
    \end{array} \right)
    \label{eq:low-energy_BdG}~,
\end{equation}
and justifies our treatment of the opposite edges as isolated edges throughout the main text. In Fig. \ref{fig:low_energy_spectrum} we can see the good agreement between the numerical results and this reduced model.

\end{document}